# On the Role of Demagnetizing Tensors in Arbitrary Orientations of General Ellipsoid: Implications for MRI Safety Assessment

Tomppa Pakarinen, Tampere Universities Faculty of Medicine and Health Technology, Tampere, Finland (e-mail: tomppa.pakarinen@tuni.fi)

**Abstract—**

**Objective**

**This work explores the behaviour of demagnetizing tensors for general ellipsoids under arbitrary rotations in homogeneous magnetic fields. The work is motivated by the concerns in magnetic resonance imaging safety and their practical evaluation in clinical environments. Whereas demagnetizing tensor is a welldefined concept in the principal axes, its transformation under three-dimensional reorientation is often overlooked a justifiable omission for solutions derived from Poisson equation, where the tensor can be directly rotated. However, this does not hold for common approximations, where such tensor is not explicitly defined.**

**Approach**

**This work demonstrates the validity of directly rotating the orthogonal basis solutions, derived from Poisson's equation, and uses the procedure to evaluate a practical approximation, based on orthogonal area-projections. The tensor rotation approach is also applied to generalize force and torque calculations for ellipsoids under three-dimensional reorientation.**

**Results**

**The results show an exact match for translation force and torque when compared to the standard single axis rotation. Additionally, a unique connection to the wellknown MRI magic angle is found as the point of convergence for prolate spheroid aspect ratios while solving the corresponding saturation field magnitudes. Finally, the evaluated approximate method was demonstrated to perform fairly across prolate and oblate spheroids. Similar approximations might extend to irregular shapes, but numerical validation would likely remain preferable due to the complexity of internal field distributions.**

**Significance**

**The presented approach offers an alternative method for force and torque calculations, such as those provided by ASTM and ISO , by generalizing the conventional trigonometric approach for spheroids. The corresponding solution to the generalized area projection approximation via Singular Value Decomposition (SVD) offers a straightforward and efficient extension to the original method and accumulates its practical utility in projection-based problems. Finally, the connection of demagnetizing tensor rotation and the magnetic resonance magic angle provides a new perspective to the phenomenon.**

## INTRODUCTION

Magnetization of an object subject to an external magnetic field depends primarily on the material's magnetic susceptibility ($\chi$), geometry and orientation of an object with respect to the external magnetic field. The induced magnetization ($\boldsymbol{M}$) gives rise to demagnetizing field ($\boldsymbol{H}_d$), opposing $\boldsymbol{M}$. In anisotropic geometries, the demagnetizing field is found via magnetostatic equations by solving the magnetic scalar potential ($\varphi$), resulting from the internal magnetization. The usual path includes solving the Poisson equation for magnetic scalar potential

$$\nabla^2\varphi(\boldsymbol{r}) = \begin{cases} \nabla\cdot\boldsymbol{M}(\boldsymbol{r}) = -\nabla\cdot\boldsymbol{H}_d(\boldsymbol{r}) = 4\pi\rho_c \\ 0 \end{cases}, \quad (1)$$

where $\rho_c$ is the charge density. The top part of (1) holds within the specimen, and the zero-condition outside of the specimen. To simplify, a uniform magnetization ($\nabla\cdot\boldsymbol{M}=0$) is often assumed, reducing the problem to Laplace's equation. The validity of the assumption depends largely on the inspected shape, material anisotropy and on the application at hand, though an important peculiarity of ellipsoidal geometries is that such definition applies in general. In addition, the solution is typically sought only for domains without free charge carriers, a fundamental condition for the demagnetizing field relation

$$\boldsymbol{H}_d(\boldsymbol{r}) = -\nabla\varphi(\boldsymbol{r}) \quad (2)$$

to be consistent with Ampere's law [1]. In the general case, Poisson equation is solved over the volume and at the surface of the object using Green's functions, the solutions for the Laplacian operator in these domains

$$\varphi(\boldsymbol{r}) = \frac{1}{4\pi}\oiiint \frac{\boldsymbol{\nabla}'\cdot\boldsymbol{M}}{|\boldsymbol{r}-\boldsymbol{r}'|}dV' + \frac{1}{4\pi}\oiint \frac{\boldsymbol{n}\cdot\boldsymbol{M}}{|\boldsymbol{r}-\boldsymbol{r}'|}dS. \quad (3)$$

Due to uniformity assumption i.e., zero-divergence within the specimen, the volume integral term vanishes, leaving the surface integral as the only contributing factor to the scalar potential. The linear relation between $\boldsymbol{H}_d$ and the corresponding magnetization is described by rank-2 demagnetizing tensor ($\mathcal{N}_{ij}$) in which the principal axes are referred to as demagnetizing factors ($N_{ii}$)

$$\boldsymbol{H}_{in}(\boldsymbol{r}) = \boldsymbol{H}_0 - \boldsymbol{H}_d = \boldsymbol{H}_0 - \mathcal{N}_{ij}\boldsymbol{M}, \tag{4}$$

where $\boldsymbol{H}_d$ and $\boldsymbol{M}$ are now reduced to vectors, $\boldsymbol{H}_0$ is the applied field and for the demagnetizing factors holds

$$\sum_i N_{ii} = 1\,. \tag{5}$$

The final tensor is found by solving the surface integral in (3) and via the relations in (2) and (4). The computation is in most cases carried out numerically since general solutions even for primitive shapes are typically incomplete, and only few regular geometries exhibit spatially constant magnetization, rendering the problem difficult for analytical solutions. Regardless, it is common to approximate the net or average demagnetizing contribution in primitive shapes using the uniformity assumption [2, 3, 4, 5], introducing some alternative approaches.

## A. Motivation and Clinical Applications

The motivation for this work, and its potential application resides in medical magnetic resonance imaging, where heating effects, translational forces and torque on foreign objects and implants are a fundamental concern [6, 7, 8] and must be evaluated individually for each situation. In majority of clinics, patient specific simulations are not accessible nor feasible, and often a swift decision is based solely on an empirical judgment. Construction of approachable methodologies with sufficient accuracy could support decision making – or at least provide intuitive insight into the governing physical phenomena. Ideally, the demagnetizing tensor can be directly applied to estimate the total forces exserted on the specimen.

## B.Problem Framing and Aim

Practical approaches to approximate demagnetizing factors on primitive geometries have been investigated extensively for decades. Even in the modern day, simplified expressions are relevant, being approachable to a broader occupational audience, and most importantly, are straightforward to implement even with basic computational tools. The past research has been focusing in the orthogonal orientations of the specimen with respect to the external magnetic field lines [2, 9, 10, 11]. However, solutions under arbitrary 3D-rotations are not necessarily trivial for such approximations, yet no prior inquiries on the topic have been published to the best of my knowledge. The evaluation is implemented here based on an area-projection method and, as such, also serves as an extension to the procedure introduced in [9]. Moreover, the presented solution in direct tensor rotation offers an alternative approach to formulate the orientational effects on translational forces and torque, also generalizing the commonly used single axis expression for spheroids to general ellipsoids under 3D-rotations, conveniently contained in a single tensor.

# METHODS

## A. Translational force

The translational force ($\boldsymbol{F}_{Trans}$) exserted on magnetizable bodies can be calculated as the gradient of the inner product between magnetization and external field vectors

$$\boldsymbol{F}_{trans} = \boldsymbol{\nabla}(\boldsymbol{B}\cdot\boldsymbol{M}), \tag{6}$$

from which it is apparent that $\boldsymbol{F}_{trans}$ is zero for the homogenous field region. The contribution of susceptibility and (scalar) demagnetizing factors can be straightforwardly derived by substituting (4) and $\boldsymbol{M} = \chi\boldsymbol{H}_{in}$ to (6)

$$\boldsymbol{F}_{trans} = \frac{\boldsymbol{F}_g}{\mu_0 g \rho_m}\frac{\boldsymbol{B}_0}{(\frac{1}{\chi}+N)}|\boldsymbol{\nabla B}|, \tag{7}$$

where $F_g$ is the gravitational force and $\rho_m$ is the mass density [12]. Consequently, in ferromagnetic materials with high susceptibility, $\boldsymbol{F}_{trans}$ scales primarily with reciprocal of $N$, which effectively acts as a shape-anisotropic contribution to magnetic susceptibility. However, (7) inherently assumes orthogonal alignment with the magnetic field lines. A trigonometric extension defines exserted translational force for spheroids under rotation as

$$\boldsymbol{F}_{trans} = \frac{V}{\mu_0}\left(\frac{\cos^2\theta}{\left(\frac{1}{\chi}+N_n\right)}+\frac{\sin^2\theta}{\left(\frac{1}{\chi}+N_\perp\right)}\right)\boldsymbol{B}_0|\boldsymbol{\nabla B}|, \tag{8}$$

where $\theta$ is referred as the angle between $\boldsymbol{H}_0$ and $\boldsymbol{M}$ due to shape anisotropy, $N_\perp$ is the demagnetizing factor perpendicular to $N_n$ and within the deflection plane, and $V$ is the object volume [13]. The expression, however, considers single axis rotation and the effect on scalar $N_{ii}$ upon reorientation and hence $\mathcal{N}$ as a tensor quantity again not explicitly addressed. Such approaches are typical also in the related research [7, 14, 15], which is fair for spheroids due to symmetry. However, this does not extend to general ellipsoids with three unique axes of symmetry.

It should be noted that at saturation magnetization in high magnetic field strengths (approximately above 1.5 $T$ in soft magnetic materials), the practical need to compute $N$ diminishes due to its negligible effect on translational forces at magnetization asymptote i.e., when the external field strength has overcome the demagnetizing effects. [12] Though shape anisotropy is still relevant in fringe fields and in modern low-field imaging systems, the omission is valid for a large proportion of clinical MRI scanners near the bore.

## B. Torque

Whereas the risks associated with translational forces consider mainly objects external to the human body, implant and foreign object torsion is a relevant risk to the nearby structures, such as arteries and neural tissue. In addition, torque even without relevant torsion tends to compromise patient comfort during the imaging procedure. Torque reaches its peak value at the highest field strength i.e., within the homogenous region of $\boldsymbol{B}_0$. In contrast, translational forces are proportional to the field gradient, highest near the bore opening and zero under ideal field homogeneity. The fundamental relation between exserted torque and a magnetic dipole ($\mathbf{m}$) is

$$\boldsymbol{T} = \boldsymbol{m} \times \boldsymbol{B}. \tag{9}$$

There is an extensive body of research around the topic, including the definitions in the International Standards Organization (ISO) and further recommendations for measuring the torque by ASTM. [7, 12, 14, 16]. A typical formulation of the torque for a single axis rotation is calculated as

$$\boldsymbol{T}_{n\times\perp} = \mu_0 V \frac{|N_n - N_\perp|}{2N_n N_\perp} |\boldsymbol{H}_0|^2 \sin 2\theta\,, \tag{10}$$

where the torque is dependent on $\boldsymbol{H}_0$ until saturation field strength ($\boldsymbol{H}_{sat}$) for which, the corollary expression is provided by Abbot et al. in [14]. Similar to (7) and (8), the basis of the torque calculation lies within the angle ($\theta$) between $\boldsymbol{H}_0$ and $\boldsymbol{M}$.

## C. Direct Tensor Rotation

Considering the dependence of demagnetizing factors solely on geometrical properties i.e., aspect ratio and orientation with respect to the external magnetic field, it may be intuitively compelling to directly rotate the orthogonal tensor instead of the shape via

$$\mathcal{N}' = \boldsymbol{R}\mathcal{N}\boldsymbol{R}^T \mid \boldsymbol{R} \in SO(3)\,, \tag{11}$$

and thus, sparing the extra work. Indeed, such intuition does apply to the exact solutions. A quantitative comparison between numerically solved Poisson equation and the direct tensor rotation is presented in Fig. 1 for a trivial case, limited only by sampling and computational accuracy, as expected.

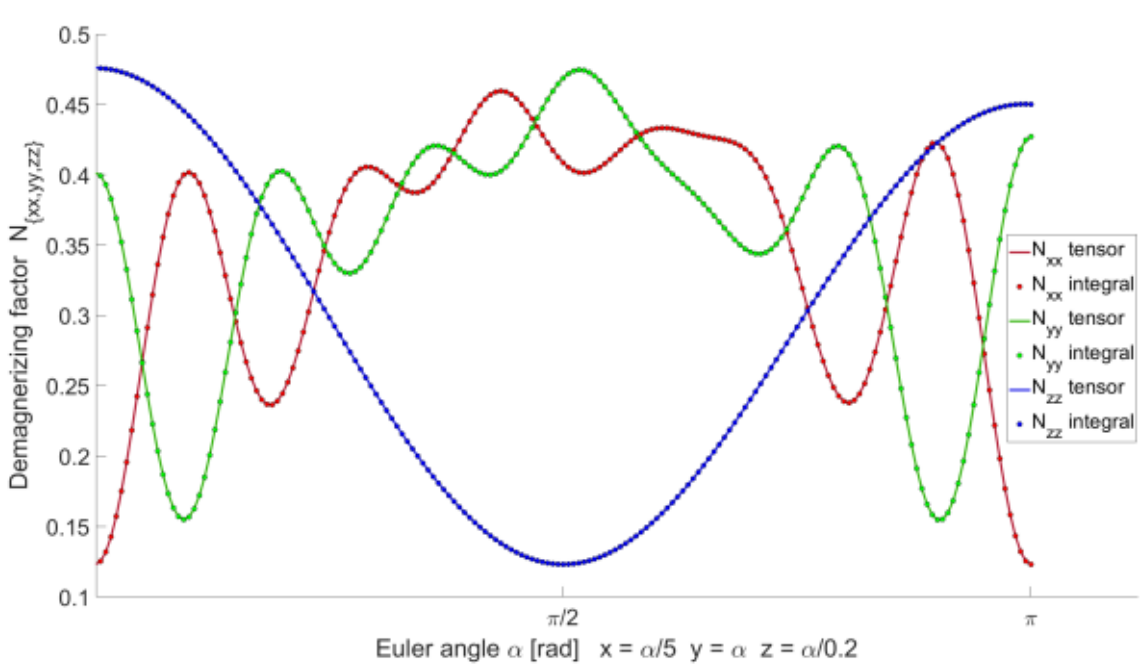

Fig. 1. *Demagnetizing factors for a general ellipsoid* ($a/b = 0.4, c/a = 0.342$) *via numerical surface integration of (3) after rotation and directly rotating the orthogonal basis solution for the demagnetizing tensor* [17].

Repeating the computation for multiple tabulated basis solutions in [17] with randomized rotations are equivalent without exceptions. $\mathcal{N}'$ can be straightforwardly applied to the general expression of translational force in (6) and torque in (9), yielding the final tensor forms as

$$\boldsymbol{F}'_{trans} = \frac{V}{\mu_0}\{(\boldsymbol{I} + \chi\mathcal{N}')^{-1}\chi\}\boldsymbol{H}_0|\boldsymbol{\nabla B}|, \tag{12}$$

and for torque

$$\boldsymbol{T}' = \mu_0 V\{(\boldsymbol{I} + \chi\mathcal{N}')^{-1}\chi\boldsymbol{H}_0\} \times \boldsymbol{H}_0, \tag{13}$$

where $\boldsymbol{I}$ is the identity matrix and $\boldsymbol{H}_0$ ranges from zero until $\boldsymbol{H}_{sat}$.

## D. Area-projection Approximation

When the inspected geometry fulfills or closely approximates conditions {$c_1$, $c_2$, $c_3$}, the average demagnetizing factors can be estimated by

$$N_{||} \approx \frac{A_\perp}{\sum A_P}, \tag{14}$$

where $A_\perp$ is the surface projection area in the perpendicular plane to the magnetic field direction and $N_{||}$ is the demagnetizing factor in the direction of $\boldsymbol{H}_0$, and $A_P$ corresponds to each orthogonal projection on a static basis [9]. Conditions $c_{1-3}$ are defined here as

$$\begin{cases} c_1: & \textit{The surface is whole and continuous} \\ c_2: & \textit{The surface geometry is convex} \\ c_3: & \textit{Geometry has axial symmetry} \end{cases}$$

Moreover, the spatial variance of material susceptibility is assumed here to have negligible effect on the overall magnetization [18, 9], allowing a solely geometrical characterization. However, directly rotating the orthogonal solutions is not applicable, since the constructed diagonal matrix is not an actual tensor and would also imply equivalence of transforms prior ($\mathbb{R}^3 \rightarrow \mathbb{R}^3$), versus after projection ($\mathbb{R}^2 \rightarrow \mathbb{R}^3$). Hence, demagnetizing factors for a given geometry must be evaluated via the orthogonal premise through $A_P$ on the static magnetic field coordinates after the re-orientation.

For a vector space $V \subset \mathbb{R}^3$, an orthogonal projection is defined as $P: V \rightarrow V$ and $P^2 = P = P^T, s.t.$

$$\boldsymbol{P}_{xy} = \begin{bmatrix} 1 & 0 & 0 \\ 0 & 1 & 0 \\ 0 & 0 & 0 \end{bmatrix}, \boldsymbol{P}_{xz} = \begin{bmatrix} 1 & 0 & 0 \\ 0 & 0 & 0 \\ 0 & 0 & 1 \end{bmatrix}, \boldsymbol{P}_{yz} = \begin{bmatrix} 0 & 0 & 0 \\ 0 & 1 & 0 \\ 0 & 0 & 1 \end{bmatrix} \quad (15)$$

Let $S \subseteq V$ be any surface, and area of orthogonal projections of rotated $S$ is defined here as

$$\begin{aligned} A_P &= Area\{\boldsymbol{x}' \mid \boldsymbol{x} \in S \wedge \boldsymbol{x}' = \boldsymbol{P}_{nk}\boldsymbol{R}(\gamma,\beta,\alpha)\boldsymbol{x}\} \\ &\equiv O_{A(\boldsymbol{u}_n^{'},\boldsymbol{u}_k^{'})_{\{n,k\}\in\{x,y,z\},n\neq k}} \end{aligned} \quad (16)$$

where $O_{A(\boldsymbol{u}'_n,\boldsymbol{u}'_k)}$ stands for a generalized 'orthogonal area operator' on the appropriate semiaxis vectors $(\boldsymbol{u}'_n, \boldsymbol{u}'_k)$, to obtain the area of the desired projection, and $\boldsymbol{R}(\gamma,\beta,\alpha) \in SO(3)$ is presented here in Euler angles. Hence, the remaining task is to define the semiaxis vectors on the orthogonal projections and the corresponding $O_{A(\boldsymbol{u}'_n,\boldsymbol{u}'_k)}$. Finally, the constraints $c_{1-3}$ ensure that for all projections, the total overlapping (obscured surface) area equals to $A_P$.

### D.1) General Ellipsoid and Area Projection

The area operator $O_{A(\boldsymbol{u}'_n,\boldsymbol{u}'_k)}$, defined in (16), for an ellipsoidal surface $S_{ell} \subseteq \mathbb{R}^3$ is simply $|\boldsymbol{u}'_n||\boldsymbol{u}'_k|\pi$, where $\boldsymbol{u}'_{\{n,k\}}$ are the semiaxis vectors for $\boldsymbol{x}' \in S'$ i.e., the projected shape on the static orthogonal planes. There are various ways finding $\boldsymbol{u}'_{\{n,k\}}$ under arbitrary rotations, but here Singular Value Decomposition (SVD) is utilized to solve the eigenvector directions and magnitudes, for which the latter relates inversely to the projected shape semiaxis lengths. Furthermore, SVD is a convenient method in obtaining the eigenvalues and the transformed basis particularly for an ellipsoidal shape, since an analogous operation is embedded within the visual interpretation of the transform itself [19]. SVD states that for any $m \times n$ matrix,

$$\boldsymbol{M} = \boldsymbol{U}\boldsymbol{\Sigma}\boldsymbol{V}^T, \quad (17)$$

where $\boldsymbol{U}$ is a left orthogonal matrix corresponding to the basis vectors of the transformed coordinates, $\boldsymbol{\Sigma}$ holds the singular values of $\boldsymbol{M}$, and $\boldsymbol{V}^T$ is the transpose of the transformation from the static basis to the output domain. The quadratic form of an ellipsoid is

$$\boldsymbol{x}^T\boldsymbol{M}\boldsymbol{x} = [x \quad y \quad z] \begin{bmatrix} \frac{1}{a^2} & 0 & 0 \\ 0 & \frac{1}{b^2} & 0 \\ 0 & 0 & \frac{1}{c^2} \end{bmatrix} \begin{bmatrix} x \\ y \\ z \end{bmatrix} = 1, \quad (18)$$

where $\mathbf{M}$ is a 'shape matrix'. Rewriting (18) by applying rotation $\boldsymbol{y} = \boldsymbol{R}\boldsymbol{x} \rightarrow \boldsymbol{x} = \boldsymbol{R}^T\boldsymbol{y}$ yields

$$(\boldsymbol{R}^T\boldsymbol{y})^T\boldsymbol{M}(\boldsymbol{R}^T\boldsymbol{y}) = \boldsymbol{y}^T\boldsymbol{R}\boldsymbol{M}\boldsymbol{R}^T\boldsymbol{y} = \boldsymbol{y}^T\boldsymbol{M}'\boldsymbol{y}. \quad (19)$$

Finally, applying projection and dimension erasing, and solving the singular values from $\boldsymbol{\Sigma}\,(\sigma_{11}, \sigma_{22})$ via SVD, expression in (16) becomes

$$A_P = \frac{\pi}{\sqrt{\sigma_{11}\sigma_{22}}} \quad (20)$$

where subindices 11 and 22 correspond to the sorted (descending) magnitudes of $\sigma$, and $\sqrt{\sigma_i}$ are the eigenvalues of collapsed $(\boldsymbol{P}_{nk}\boldsymbol{M}'^{-1})^{-1}$. It is worth noting that for spheroids, only $\sigma_{22}$ is of interest, since $\frac{1}{\sqrt{\sigma_{11}}} = |u_{minor}|$ under all rotations.

### D.2) Ensuring SVD Equivalence

A quantitative validation of the method is straightforward by demonstrating the correspondence of SVD with a numerical solution of a point cloud, enclosed by an ellipsoid surface

$$\frac{x^2}{a^2} + \frac{y^2}{b^2} + \frac{z^2}{c^2} = 1. \quad (21)$$

Finding the area of the projected convex hull on the orthogonal static planes yields the same result as with SVD, limited only by sampling and computational accuracy, as expected.

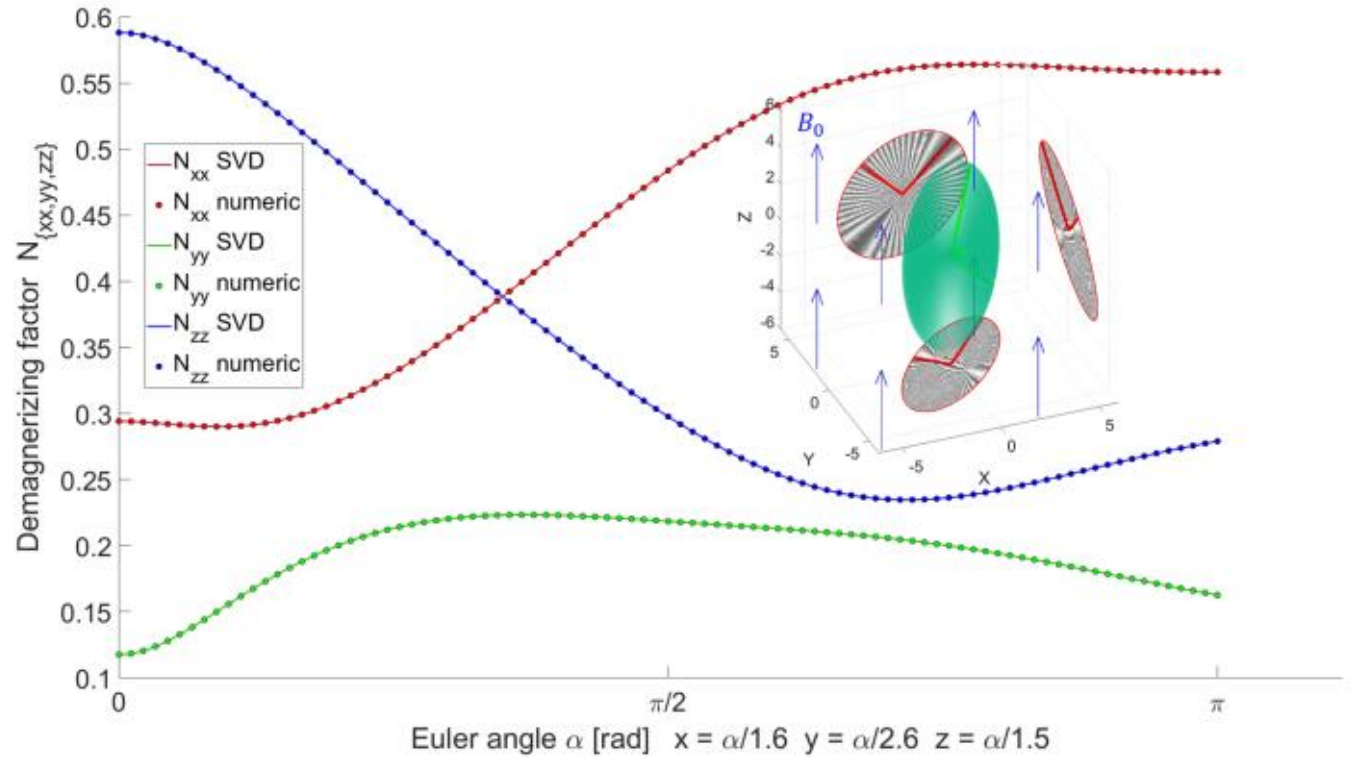

Fig. 2. *Example of demagnetizing factors of a general ellipsoid* ($a = 2,\ b = 5,\ c = 1$) under arbitrary rotations *via Singular Value Decomposition (SVD) and via numerical solution of the convex hull area. The point cloud projections are illustrated within the embedded figure together with corresponding SVD solutions, presented as red-colored semiaxes.*

The accuracy of the area-projection approximation itself is most appropriately evaluated using the exact analytical solutions of (3) - (5), which can now be carried out conveniently by comparing SVD and the direct tensor rotation.

# RESULTS

## A. Model Comparison – Accuracy of the Area-projection Model in Spheroids

Exact solutions for the orthogonal demagnetizing factors were utilized from [17] to construct a reference via direct tensor rotation. A quantitative analysis for the area-projection approximation is presented in Fig. 3.

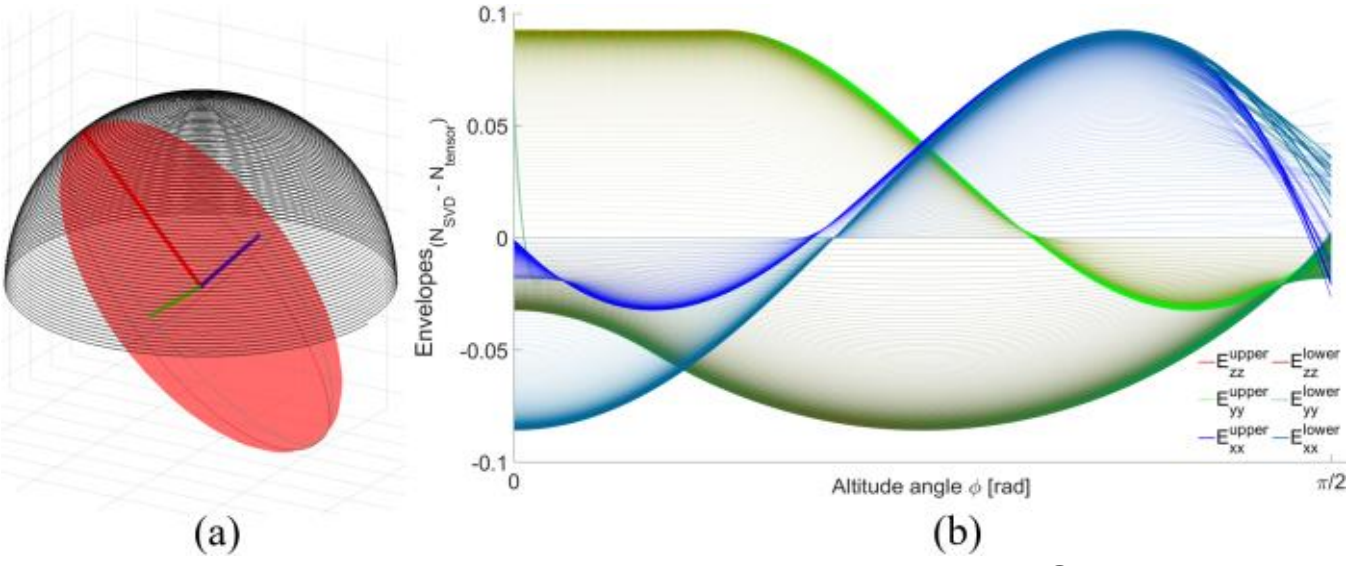

Fig. 3. *The area-projection method using Singular value decomposition versus directly rotating the demagnetizing tensor for prolate spheroids with various aspect ratios. The orthogonal solutions were computed using the expression in* [17]. *The sampling function follows a hemispherical spiral path with 0.005 rad ascent per revolution, illustrated in (a). The difference between the models is presented as upper and lower envelopes ($E^{u,l}_{nn}$). Each envelope pair encompasses an alternating curve (due to azimuthal rotation, not visualized) for different aspect ratios ($b/a = \{0.999 - 0.0067\}, b = c$). Higher line density corresponds to a higher elongation of the spheroid.*

Here each envelope corresponds in practice to the maximum absolute error during azimuthal revolution for the given aspect ratio. Absolute error was chosen over the relative error due to latter's inflated values, disproportionate to their physical significance when the inspected demagnetizing component is small, also acknowledged in [9]. Finally, the model performance was evaluated as the function of aspect ratio over the sampled orientations via coefficient of determination $R^2$.

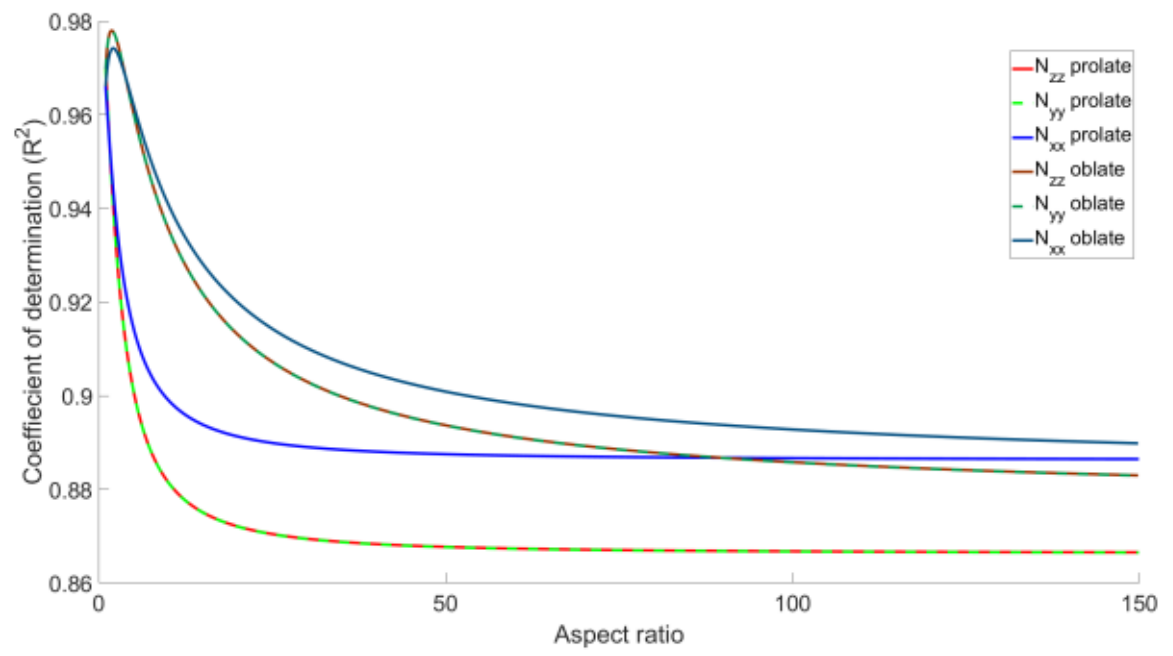

Fig. 4. *Model performance for SVD method in approximating the demagnetizing tensor under rotations ($\theta = 0 - 2\pi, \varphi = 0 - \frac{\pi}{4}$) measured with the coefficient of determination ($R^2$) as the function of aspect ratio.*

Note that Osborn's formulas for prolate and oblate spheroids are undefined at $a/b = 1$ due to singularity. For the spherical case, the demagnetizing factors reduce to $N_{xy} = N_{xz} = N_{yz} = \frac{1}{3}$, which must be treated separately. The disparity is apparent in Fig. 4 when $a/b \rightarrow 1$, for which $R^2 = 1$ would be expected.

## B. Translational force and Torque

As shown in (12) and (13), the direct consequence of the tensor rotation method is the reformulation of scalar expressions in (7) and (10), also generalizing them for general ellipsoids under arbitrary orientations. The comparison of the two expressions with respect to the standard trigonometric formulation in spheroids for single axis rotations is presented in Fig. 5.

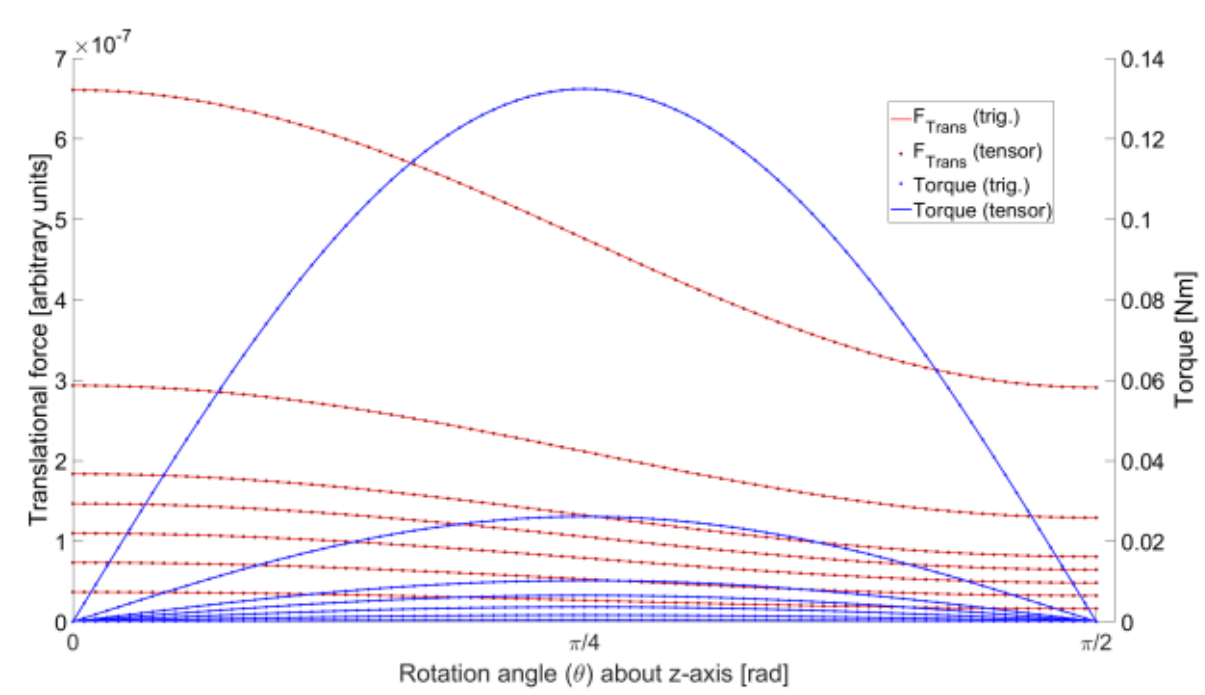

Fig. 5. Translational force and torque comparison to expression in (7) and (9). Corresponding physical constants and field strengths were matched with [14]. The force was computed here with setting the field gradient strength to unity.

The results show the generalized solutions for torque and translational force approach the trigonometric single axis method for $\chi \gg 1$, as expected.

## D. Saturation Field Magnitude

Since torque and translational forces are dependent on the absolute field strength until saturation, it is sensible to recompute the angle dependent saturation field magnitudes. The comparison was performed against the expression in Abbot et al. for several aspect ratios, presented in Fig. 6.

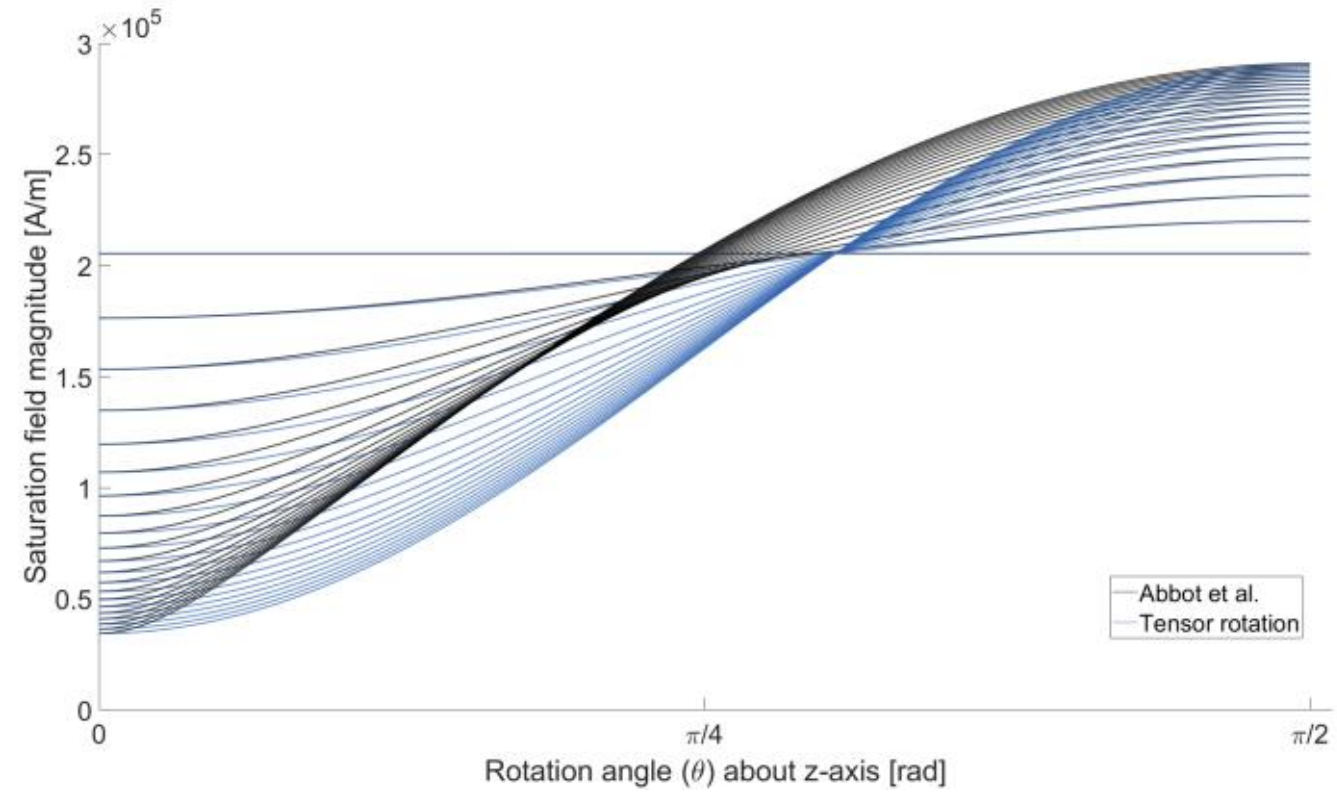

Fig. 6. Saturation field strength for prolate spheroid, computed using the tensor rotation method against the expression in Abbot et al. The results differ in non-orthogonal orientations by increasing magnitude with higher aspect ratios. Note that the tensor method solutions convergence to a single point at $0.955317\ rad = 54.7356º$.

The most notable caveat in Fig 6. is not the absolute difference between the approaches, but the convergence of tensor rotation solutions to a single point over all aspect ratios at a rather interesting angle of $54.7356º$, commonly known as the MRI magic angle.

## DISCUSSION AND CONCLUSIONS

Demagnetizing tensor determination and their practical approximations have important applications in magnet manufacturing, computational modelling, magnetic manipulation, and most relevantly, in magnetic resonance imaging, particularly in medical contexts where magnetic safety is a major concern. The presence of foreign objects within the body is a common contraindication for MRI – primarily due to the induced eddy currents and heating, but also due to the translational forces and torque inflicted by the strong magnetic field. This work provides means to further approximate the latter two under arbitrary orientations, with minimal effort.

The main contribution of this work can be summarized as examining the details of object rotation effect on demagnetizing tensors with respect to:

1. Generalized solutions using simple tensor rotation
2. Area-projection approximation [9] and its applicability under arbitrary rotations
3. Generalized tensor-based formulation for translational force and torque calculations.
4. Connecting demagnetizing tensor and the saturation magnetic field magnitudes to the well-known magic angle.

The first point is a direct consequence of the geometric interpretation of demagnetizing tensors, simplifying the process of finding an exact solution under rotations, given that a single base solution is known. The approach is seemingly unnoticed in MRI research, though the presented formulation aligns with similar approaches that have appeared in other contexts, including early theoretical work by Della Torre [20] and recent applications in quantum magnomechanics [21]. The method may be directly applied in generalized force, torque and saturation field calculations, thus leading in point 3 and 4. In point 4, the manifestation of the magic angle as the converging point of prolate spheroid aspect ratios is an interesting observation − effectively, the diagonal demagnetizing factors and the saturation magnetization become independent of the shape anisotropy at this very specific orientation. The magic angle ($\vartheta$) is traditionally derived from average dipolar interactions, formally described via a second order Legendre polynomial, for which the roots are $\pi + (-1)^n(\pi - \vartheta), n \in \mathbb{N}_1$, corresponding exactly to the periodic convergence points upon tensor rotation.

In point 2, a general non-applicability of the area-projection method is evident when considering any geometry, not subject to $c_{1-3}$. For example, an arbitrary projection of a concave surface does not ensure 1: 1 linear mapping on to the obscured segment. For the applicable cases, the major pitfall boils down to the definition of $O_{A(\boldsymbol{u}'_n,\boldsymbol{u}'_k)}$, which must be constructed individually for each geometric shape, and to the fact that even for primitive geometries, only few exhibit spatially independent magnetization. However, once a connection is constructed and underlying approximations are accepted, it will be applicable for all aspect ratios. In addition, SVD proves to be useful for the case of general ellipsoid due to its quadratic form, but only when the transformed basis vector directions are of interest. Otherwise, simpler closed-form expressions may be more suitable. For non-quadratic shapes, one might consider computing an equivalent ellipsoid [22], solving the problem via trigonometry, or finding a solution to the point cloud similar to Fig. 1.

The comparative results suggest that the area-projection method is applicable in approximating demagnetizing factors for spheroids under arbitrary 3D rotations with fair accuracy ($0.86 < R^2 < 0.97$) − comparable to their trivial cases in orthogonal orientations. Although the current analysis is limited to quantitative results for prolate and oblate spheroids with finite sampling, the presented method and equations are constructed for a general ellipsoid and are straightforward to reproduce for independent validation.

It may be concluded that, when a solution to the Poisson equation exists in an orthogonal configuration, one should consider directly transforming the corresponding demagnetizing tensor. Moreover, approximations derived directly from axial lengths [9, 2] may not translate accurately to irregular geometries, not subject to $c_{1-3}$; in such cases, numerical solutions remain preferable, though the topic does carry some academic appeal. Accordingly, the method's practical value lies primarily in shapes, that meet or closely approximate these conditions.

## ACKNOWLEDGMENT

This research received no specific grant from any funding agency in the public, commercial, or not-for-profit sectors.

## REFERENCES

[1] D. Griffiths and R. College, Introduction to Electrodynamics, New Jersey: Prentice Hall, 1999.

[2] M. Sato and Y. Ishii, "Simple and approximate expressions of demagnetizing factors of uniformly magnetized rectangular rod and cylinder," *Journal of Applied Physics,* vol. 66, no. 2, pp. 983-985, 1989.

[3] T. Taniguchi, "An analytical computation of magnetic field generated from a cylinder ferromagnet," *Journal of the Physical Society of Japan,* vol. 87, no. 4, p. 044802, 2018.

[4] F. Lang and S. Blundell, "Fourier space derivation of the demagnetization tensor for uniformly magnetized objects of cylindrical symmetry," *Journal of Magnetism and Magnetic Materials,* vol. 401, pp. 1060-1067, 2016.

[5] D. Chen, E. Pardo and A. Sanchez, "Demagnetizing Factors for Rectangular Prisms," *IEEE Transactions on Magnetics,* pp. 2077-2088, 2005.

[6] G. Durbridge, "Magnetic resonance imaging: fundamental safety issues.," *Journal of orthopaedic & sports physical therapy,* vol. 41, no. 11, pp. 820-828, 2011.

[7] i. ASTM, "Standard Test Method for Measurement of Magnetically Induced Displacement Force on Medical Devices in the Magnetic Resonance Environment.," ASTM International, West Conshohocken, PA, 2021.

[8] I. C. o. N.-I. R. P. ICNIRP, "ICNIRP Statement—Guidelines on Limits of Exposure to Static Magnetic Fields," *Health Physics,* vol. 96, no. 4, pp. 504-514, 2009.

[9] C. Bahl, "Estimating the demagnetization factors for regular permanent magnet pieces," *AIP Advances,* no. 11, p. 075028, 2021.

[10] B. Pugh, D. Kramer and C. Chen, "Demagnetizing Factors for Various Geometries Precisely Determined Using 3-D Electromagnetic Field Simulation," *IEEE Transactions on Magnetics,* vol. 47, pp. 4100-4103, 2011.

[11] K. Etse and X. Mininger, "Determination of demagnetizing factors for various geometries using an iterative numerical approach," *Journal of Magnetism and Magnetic Materials,* vol. 564, no. 2, p. 170151, 2022.

[12] L. Panych and P. Madore, "The physics of MRI safety.," *Magn Reson Imaging.,* pp. 28-43, 2018.

[13] J. Schenck, "Safety of Strong, Static Magnetic Fields," *Journal of Magnetic Resonance Imaging,* vol. 12, no. 1, pp. 2-19, 2000.

[14] J. Abbot, O. Ergeneman, M. Kummer, A. Hirt and B. Nelson, "Modeling Magnetic Torque and Force for Controlled," *IEEE TRANSACTIONS ON ROBOTICS,* vol. 23, no. 6, pp. 1247-1252, 2007.

[15] P. Jackson, "Dancing paperclips and the geometric influence on magnetization:," *American Journal of Physics,* vol. 74, no. 4, pp. 272-279, 2006.

[16] International Organization for Standardization, "Assessment of the safety of magnetic resonance imaging for patients with an active implantable medical device — Part 1: Requirements for information before magnetic resonance imaging," ISO, 2018.

[17] J. Osborn, "Demagnetizing factors of the general ellipsoid.," *Physical Review,* vol. 67, no. 11-12, pp. 351-357, 1945.

[18] R. Prozorov and V. Kogan, "Effective Demagnetizing Factors of Diamagnetic Samples of Various Shapes," *Physical Review Applied,* vol. 1, no. 10, p. 014030, 2017.

[19] J. Raghavendar and V. Dharmaiah, "Geometrical Interpretation of Singular Value Decomposition (SVD) & Applications of SVD," *International Journal of Scientific and Innovative Mathematical Research,* vol. 5, no. 4, pp. 23-26, 2017.

[20] E. Della Torre, "Theoretical Apects of Demagnetizing Tensors," *IEEE Transactions on Magnetics,* vol. 2, pp. 739-744, 1967.

[21] N. Bort-Soldevila, J. Cunill-Subiranas, N. Del-Valle, W. Wieczored, G. Higgins, M. Trupke and C. Navau, "Modeling magnetically levitated superconducting ellipsoids, cylinders, and cuboids," *PHYSICAL REVIEW RESEARCH 6,* vol. 6, no. 4, p. 043046, 2024.

[22] M. Beleggia, M. Graef and M. Yonko, "The equivalent ellipsoid of a magnetized body," *Journal of Physics D: Applied Physics,* vol. 39, no. 5, pp. 891-899, 2006.

[23] A. Ferreira, R. Siskey and J. White, "Magnetically Induced Displacement Force on Medical Devices," in *Proceedings of the 2016 COMSOL Conference*, Boston, USA, 2016.